\documentclass[11pt]{article}
\usepackage{amsmath}
\usepackage{amssymb}

\def\PsfigVersion{1.9}
\ifx\undefined\psfig\else\endinput\fi

\let\LaTeXAtSign=\@
\let\@=\relax
\edef\psfigRestoreAt{\catcode`\@=\number\catcode`@\relax}
\catcode`\@=11\relax
\newwrite\@unused
\def\ps@typeout#1{{\let\protect\string\immediate\write\@unused{#1}}}
\ps@typeout{psfig/tex \PsfigVersion}

\def\figurepath{./}

\def\@nnil{\@nil}
\def\@empty{}
\def\@psdonoop#1\@@#2#3{}
\def\@psdo#1:=#2\do#3{\edef\@psdotmp{#2}\ifx\@psdotmp\@empty \else
    \expandafter\@psdoloop#2,\@nil,\@nil\@@#1{#3}\fi}
\def\@psdoloop#1,#2,#3\@@#4#5{\def#4{#1}\ifx #4\@nnil \else
       #5\def#4{#2}\ifx #4\@nnil \else#5\@ipsdoloop #3\@@#4{#5}\fi\fi}
\def\@ipsdoloop#1,#2\@@#3#4{\def#3{#1}\ifx #3\@nnil 
       \let\@nextwhile=\@psdonoop \else
      #4\relax\let\@nextwhile=\@ipsdoloop\fi\@nextwhile#2\@@#3{#4}}
\def\@tpsdo#1:=#2\do#3{\xdef\@psdotmp{#2}\ifx\@psdotmp\@empty \else
    \@tpsdoloop#2\@nil\@nil\@@#1{#3}\fi}
\def\@tpsdoloop#1#2\@@#3#4{\def#3{#1}\ifx #3\@nnil 
       \let\@nextwhile=\@psdonoop \else
      #4\relax\let\@nextwhile=\@tpsdoloop\fi\@nextwhile#2\@@#3{#4}}
\ifx\undefined\fbox
\newdimen\fboxrule
\newdimen\fboxsep
\newdimen\ps@tempdima
\newbox\ps@tempboxa
\long\def\fbox#1{\leavevmode\setbox\ps@tempboxa\hbox{#1}\ps@tempdima\fboxrule
    \advance\ps@tempdima \fboxsep \advance\ps@tempdima \dp\ps@tempboxa
   \hbox{\lower \ps@tempdima\hbox
  {\vbox{\hrule height \fboxrule
          \hbox{\vrule width \fboxrule \hskip\fboxsep
          \vbox{\vskip\fboxsep \box\ps@tempboxa\vskip\fboxsep}\hskip 
                 \fboxsep\vrule width \fboxrule}
                 \hrule height \fboxrule}}}}
\fi
\newread\ps@stream
\newif\ifnot@eof       %
\newif\if@noisy        %
\newif\if@atend        %
\newif\if@psfile       %
\chardef\psletter = 11 %
\chardef\other = 12

\newif \ifdebug %
\newif\ifc@mpute %
\c@mputetrue %

\let\then = \relax
\def\r@dian{pt }
\let\r@dians = \r@dian
\let\dimensionless@nit = \r@dian
\let\dimensionless@nits = \dimensionless@nit
\def\internal@nit{sp }
\let\internal@nits = \internal@nit
\newif\ifstillc@nverging
\def \Mess@ge #1{\ifdebug \then \message {#1} \fi}

{ %
        \catcode `\@ = \psletter
        \gdef \nodimen {\expandafter \n@dimen \the \dimen}
        \gdef \term #1 #2 #3%
               {\edef \t@ {\the #1}%
                \edef \t@@ {\expandafter \n@dimen \the #2\r@dian}%
                \t@rm {\t@} {\t@@} {#3}%
               }
        \gdef \t@rm #1 #2 #3%
               {{%
                \count 0 = 0
                \dimen 0 = 1 \dimensionless@nit
                \dimen 2 = #2\relax
                \Mess@ge {Calculating term #1 of \nodimen 2}%
                \loop
                \ifnum  \count 0 < #1
                \then   \advance \count 0 by 1
                        \Mess@ge {Iteration \the \count 0 \space}%
                        \Multiply \dimen 0 by {\dimen 2}%
                        \Mess@ge {After multiplication, term = \nodimen 0}%
                        \Divide \dimen 0 by {\count 0}%
                        \Mess@ge {After division, term = \nodimen 0}%
                \repeat
                \Mess@ge {Final value for term #1 of 
                                \nodimen 2 \space is \nodimen 0}%
                \xdef \Term {#3 = \nodimen 0 \r@dians}%
                \aftergroup \Term
               }}
        \catcode `\p = \other
        \catcode `\t = \other
        \gdef \n@dimen #1pt{#1} %
}

\def \Divide #1by #2{\divide #1 by #2} %

\def \Multiply #1by #2%
       {{%
        \count 0 = #1\relax
        \count 2 = #2\relax
        \count 4 = 65536
        \Mess@ge {Before scaling, count 0 = \the \count 0 \space and
                        count 2 = \the \count 2}%
        \ifnum  \count 0 > 32767 %
        \then   \divide \count 0 by 4
                \divide \count 4 by 4
        \else   \ifnum  \count 0 < -32767
                \then   \divide \count 0 by 4
                        \divide \count 4 by 4
                \else
                \fi
        \fi
        \ifnum  \count 2 > 32767 %
        \then   \divide \count 2 by 4
                \divide \count 4 by 4
        \else   \ifnum  \count 2 < -32767
                \then   \divide \count 2 by 4
                        \divide \count 4 by 4
                \else
                \fi
        \fi
        \multiply \count 0 by \count 2
        \divide \count 0 by \count 4
        \xdef \product {#1 = \the \count 0 \internal@nits}%
        \aftergroup \product
       }}

\def\r@duce{\ifdim\dimen0 > 90\r@dian \then   %
                \multiply\dimen0 by -1
                \advance\dimen0 by 180\r@dian
                \r@duce
            \else \ifdim\dimen0 < -90\r@dian \then  %
                \advance\dimen0 by 360\r@dian
                \r@duce
                \fi
            \fi}

\def\Sine#1%
       {{%
        \dimen 0 = #1 \r@dian
        \r@duce
        \ifdim\dimen0 = -90\r@dian \then
           \dimen4 = -1\r@dian
           \c@mputefalse
        \fi
        \ifdim\dimen0 = 90\r@dian \then
           \dimen4 = 1\r@dian
           \c@mputefalse
        \fi
        \ifdim\dimen0 = 0\r@dian \then
           \dimen4 = 0\r@dian
           \c@mputefalse
        \fi
        \ifc@mpute \then
                \divide\dimen0 by 180
                \dimen0=3.141592654\dimen0
                \dimen 2 = 3.1415926535897963\r@dian %
                \divide\dimen 2 by 2 %
                \Mess@ge {Sin: calculating Sin of \nodimen 0}%
                \count 0 = 1 %
                \dimen 2 = 1 \r@dian %
                \dimen 4 = 0 \r@dian %
                \loop
                        \ifnum  \dimen 2 = 0 %
                        \then   \stillc@nvergingfalse 
                        \else   \stillc@nvergingtrue
                        \fi
                        \ifstillc@nverging %
                        \then   \term {\count 0} {\dimen 0} {\dimen 2}%
                                \advance \count 0 by 2
                                \count 2 = \count 0
                                \divide \count 2 by 2
                                \ifodd  \count 2 %
                                \then   \advance \dimen 4 by \dimen 2
                                \else   \advance \dimen 4 by -\dimen 2
                                \fi
                \repeat
        \fi             
                        \xdef \sine {\nodimen 4}%
       }}

\def\Cosine#1{\ifx\sine\UnDefined\edef\Savesine{\relax}\else
                             \edef\Savesine{\sine}\fi
        {\dimen0=#1\r@dian\advance\dimen0 by 90\r@dian
         \Sine{\nodimen 0}
         \xdef\cosine{\sine}
         \xdef\sine{\Savesine}}}              

\def\psdraft{
        \def\@psdraft{0}
}
\def\psfull{
        \def\@psdraft{100}
}

\psfull

\newif\if@scalefirst
\def\psscalefirst{\@scalefirsttrue}
\def\psrotatefirst{\@scalefirstfalse}
\psrotatefirst

\newif\if@draftbox
\def\psnodraftbox{
        \@draftboxfalse
}
\def\psdraftbox{
        \@draftboxtrue
}
\@draftboxtrue

\newif\if@prologfile
\newif\if@postlogfile
\def\pssilent{
        \@noisyfalse
}
\def\psnoisy{
        \@noisytrue
}
\psnoisy
\newif\if@bbllx
\newif\if@bblly
\newif\if@bburx
\newif\if@bbury
\newif\if@height
\newif\if@width
\newif\if@rheight
\newif\if@rwidth
\newif\if@angle
\newif\if@clip
\newif\if@verbose
\def\@p@@sclip#1{\@cliptrue}

\newif\if@decmpr

\def\@p@@sfigure#1{\def\@p@sfile{null}\def\@p@sbbfile{null}
                \openin1=#1.bb
                \ifeof1\closein1
                        \openin1=\figurepath#1.bb
                        \ifeof1\closein1
                                \openin1=#1
                                \ifeof1\closein1%
                                       \openin1=\figurepath#1
                                        \ifeof1
                                           \ps@typeout{Error, File #1 not found}
                                                \if@bbllx\if@bblly
                                                \if@bburx\if@bbury
                                                        \def\@p@sfile{#1}%
                                                        \def\@p@sbbfile{#1}%
                                                        \@decmprfalse
                                                \fi\fi\fi\fi
                                        \else\closein1
                                                \def\@p@sfile{\figurepath#1}%
                                                \def\@p@sbbfile{\figurepath#1}%
                                                \@decmprfalse
                                        \fi%
                                \else\closein1%
                                        \def\@p@sfile{#1}
                                        \def\@p@sbbfile{#1}
                                        \@decmprfalse
                                \fi
                        \else
                                \def\@p@sfile{\figurepath#1}
                                \def\@p@sbbfile{\figurepath#1.bb}
                                \@decmprtrue
                        \fi
                \else
                        \def\@p@sfile{#1}
                        \def\@p@sbbfile{#1.bb}
                        \@decmprtrue
                \fi}

\def\@p@@sfile#1{\@p@@sfigure{#1}}

\def\@p@@sbbllx#1{
                \@bbllxtrue
                \dimen100=#1
                \edef\@p@sbbllx{\number\dimen100}
}
\def\@p@@sbblly#1{
                \@bbllytrue
                \dimen100=#1
                \edef\@p@sbblly{\number\dimen100}
}
\def\@p@@sbburx#1{
                \@bburxtrue
                \dimen100=#1
                \edef\@p@sbburx{\number\dimen100}
}
\def\@p@@sbbury#1{
                \@bburytrue
                \dimen100=#1
                \edef\@p@sbbury{\number\dimen100}
}
\def\@p@@sheight#1{
                \@heighttrue
                \dimen100=#1
                \edef\@p@sheight{\number\dimen100}
}
\def\@p@@swidth#1{
                \@widthtrue
                \dimen100=#1
                \edef\@p@swidth{\number\dimen100}
}
\def\@p@@srheight#1{
                \@rheighttrue
                \dimen100=#1
                \edef\@p@srheight{\number\dimen100}
}
\def\@p@@srwidth#1{
                \@rwidthtrue
                \dimen100=#1
                \edef\@p@srwidth{\number\dimen100}
}
\def\@p@@sangle#1{
                \@angletrue
                \edef\@p@sangle{#1} %
}
\def\@p@@ssilent#1{ 
                \@verbosefalse
}
\def\@p@@sprolog#1{\@prologfiletrue\def\@prologfileval{#1}}
\def\@p@@spostlog#1{\@postlogfiletrue\def\@postlogfileval{#1}}
\def\@cs@name#1{\csname #1\endcsname}
\def\@setparms#1=#2,{\@cs@name{@p@@s#1}{#2}}
\def\ps@init@parms{
                \@bbllxfalse \@bbllyfalse
                \@bburxfalse \@bburyfalse
                \@heightfalse \@widthfalse
                \@rheightfalse \@rwidthfalse
                \def\@p@sbbllx{}\def\@p@sbblly{}
                \def\@p@sbburx{}\def\@p@sbbury{}
                \def\@p@sheight{}\def\@p@swidth{}
                \def\@p@srheight{}\def\@p@srwidth{}
                \def\@p@sangle{0}
                \def\@p@sfile{} \def\@p@sbbfile{}
                \def\@p@scost{10}
                \def\@sc{}
                \@prologfilefalse
                \@postlogfilefalse
                \@clipfalse
                \if@noisy
                        \@verbosetrue
                \else
                        \@verbosefalse
                \fi
}
\def\parse@ps@parms#1{
                \@psdo\@psfiga:=#1\do
                   {\expandafter\@setparms\@psfiga,}}
\newif\ifno@bb
\def\bb@missing{
        \if@verbose{
                \ps@typeout{psfig: searching \@p@sbbfile \space  for bounding box}
        }\fi
        \no@bbtrue
        \epsf@getbb{\@p@sbbfile}
        \ifno@bb \else \bb@cull\epsf@llx\epsf@lly\epsf@urx\epsf@ury\fi
}       
\def\bb@cull#1#2#3#4{
        \dimen100=#1 bp\edef\@p@sbbllx{\number\dimen100}
        \dimen100=#2 bp\edef\@p@sbblly{\number\dimen100}
        \dimen100=#3 bp\edef\@p@sbburx{\number\dimen100}
        \dimen100=#4 bp\edef\@p@sbbury{\number\dimen100}
        \no@bbfalse
}
\newdimen\p@intvaluex
\newdimen\p@intvaluey
\def\rotate@#1#2{{\dimen0=#1 sp\dimen1=#2 sp
                  \global\p@intvaluex=\cosine\dimen0
                  \dimen3=\sine\dimen1
                  \global\advance\p@intvaluex by -\dimen3
                  \global\p@intvaluey=\sine\dimen0
                  \dimen3=\cosine\dimen1
                  \global\advance\p@intvaluey by \dimen3
                  }}
\def\compute@bb{
                \no@bbfalse
                \if@bbllx \else \no@bbtrue \fi
                \if@bblly \else \no@bbtrue \fi
                \if@bburx \else \no@bbtrue \fi
                \if@bbury \else \no@bbtrue \fi
                \ifno@bb \bb@missing \fi
                \ifno@bb \ps@typeout{FATAL ERROR: no bb supplied or found}
                        \no-bb-error
                \fi
                \count203=\@p@sbburx
                \count204=\@p@sbbury
                \advance\count203 by -\@p@sbbllx
                \advance\count204 by -\@p@sbblly
                \edef\ps@bbw{\number\count203}
                \edef\ps@bbh{\number\count204}
                \if@angle 
                        \Sine{\@p@sangle}\Cosine{\@p@sangle}
                        {\dimen100=\maxdimen\xdef\r@p@sbbllx{\number\dimen100}
                                            \xdef\r@p@sbblly{\number\dimen100}
                                            \xdef\r@p@sbburx{-\number\dimen100}
                                            \xdef\r@p@sbbury{-\number\dimen100}}
                        \def\minmaxtest{
                           \ifnum\number\p@intvaluex<\r@p@sbbllx
                              \xdef\r@p@sbbllx{\number\p@intvaluex}\fi
                           \ifnum\number\p@intvaluex>\r@p@sbburx
                              \xdef\r@p@sbburx{\number\p@intvaluex}\fi
                           \ifnum\number\p@intvaluey<\r@p@sbblly
                              \xdef\r@p@sbblly{\number\p@intvaluey}\fi
                           \ifnum\number\p@intvaluey>\r@p@sbbury
                              \xdef\r@p@sbbury{\number\p@intvaluey}\fi
                           }
                        \rotate@{\@p@sbbllx}{\@p@sbblly}
                        \minmaxtest
                        \rotate@{\@p@sbbllx}{\@p@sbbury}
                        \minmaxtest
                        \rotate@{\@p@sbburx}{\@p@sbblly}
                        \minmaxtest
                        \rotate@{\@p@sbburx}{\@p@sbbury}
                        \minmaxtest
                        \edef\@p@sbbllx{\r@p@sbbllx}\edef\@p@sbblly{\r@p@sbblly}
                        \edef\@p@sbburx{\r@p@sbburx}\edef\@p@sbbury{\r@p@sbbury}
                \fi
                \count203=\@p@sbburx
                \count204=\@p@sbbury
                \advance\count203 by -\@p@sbbllx
                \advance\count204 by -\@p@sbblly
                \edef\@bbw{\number\count203}
                \edef\@bbh{\number\count204}
}
\def\in@hundreds#1#2#3{\count240=#2 \count241=#3
                     \count100=\count240        %
                     \divide\count100 by \count241
                     \count101=\count100
                     \multiply\count101 by \count241
                     \advance\count240 by -\count101
                     \multiply\count240 by 10
                     \count101=\count240        %
                     \divide\count101 by \count241
                     \count102=\count101
                     \multiply\count102 by \count241
                     \advance\count240 by -\count102
                     \multiply\count240 by 10
                     \count102=\count240        %
                     \divide\count102 by \count241
                     \count200=#1\count205=0
                     \count201=\count200
                        \multiply\count201 by \count100
                        \advance\count205 by \count201
                     \count201=\count200
                        \divide\count201 by 10
                        \multiply\count201 by \count101
                        \advance\count205 by \count201
                     \count201=\count200
                        \divide\count201 by 100
                        \multiply\count201 by \count102
                        \advance\count205 by \count201
                     \edef\@result{\number\count205}
}
\def\compute@wfromh{
                \in@hundreds{\@p@sheight}{\@bbw}{\@bbh}
                \edef\@p@swidth{\@result}
}
\def\compute@hfromw{
                \in@hundreds{\@p@swidth}{\@bbh}{\@bbw}
                \edef\@p@sheight{\@result}
}
\def\compute@handw{
                \if@height 
                        \if@width
                        \else
                                \compute@wfromh
                        \fi
                \else 
                        \if@width
                                \compute@hfromw
                        \else
                                \edef\@p@sheight{\@bbh}
                                \edef\@p@swidth{\@bbw}
                        \fi
                \fi
}
\def\compute@resv{
                \if@rheight \else \edef\@p@srheight{\@p@sheight} \fi
                \if@rwidth \else \edef\@p@srwidth{\@p@swidth} \fi
}
\def\compute@sizes{
        \compute@bb
        \if@scalefirst\if@angle
        \if@width
           \in@hundreds{\@p@swidth}{\@bbw}{\ps@bbw}
           \edef\@p@swidth{\@result}
        \fi
        \if@height
           \in@hundreds{\@p@sheight}{\@bbh}{\ps@bbh}
           \edef\@p@sheight{\@result}
        \fi
        \fi\fi
        \compute@handw
        \compute@resv}

\def\psfig#1{\vbox {
        \ps@init@parms
        \parse@ps@parms{#1}
        \compute@sizes
        \ifnum\@p@scost<\@psdraft{
                \special{ps::[begin]    \@p@swidth \space \@p@sheight \space
                                \@p@sbbllx \space \@p@sbblly \space
                                \@p@sbburx \space \@p@sbbury \space
                                startTexFig \space }
                \if@angle
                        \special {ps:: \@p@sangle \space rotate \space} 
                \fi
                \if@clip{
                        \if@verbose{
                                \ps@typeout{(clip)}
                        }\fi
                        \special{ps:: doclip \space }
                }\fi
                \if@prologfile
                    \special{ps: plotfile \@prologfileval \space } \fi
                \if@decmpr{
                        \if@verbose{
                                \ps@typeout{psfig: including \@p@sfile.Z \space }
                        }\fi
                        \special{ps: plotfile "`zcat \@p@sfile.Z" \space }
                }\else{
                        \if@verbose{
                                \ps@typeout{psfig: including \@p@sfile \space }
                        }\fi
                        \special{ps: plotfile \@p@sfile \space }
                }\fi
                \if@postlogfile
                    \special{ps: plotfile \@postlogfileval \space } \fi
                \special{ps::[end] endTexFig \space }
                \vbox to \@p@srheight sp{
                        \hbox to \@p@srwidth sp{
                                \hss
                        }
                \vss
                }
        }\else{
                \if@draftbox{           
                        \hbox{\frame{\vbox to \@p@srheight sp{
                        \vss
                        \hbox to \@p@srwidth sp{ \hss \@p@sfile \hss }
                        \vss
                        }}}
                }\else{
                        \vbox to \@p@srheight sp{
                        \vss
                        \hbox to \@p@srwidth sp{\hss}
                        \vss
                        }
                }\fi

        }\fi
}}
\psfigRestoreAt
\let\@=\LaTeXAtSign

\makeatletter%
\def\nottoobig#1{{\hbox{$\left#1\vcenter to1.111\ht\strutbox{}\right.\n@space$}}}
\makeatother%

\makeatletter%
\def\@begintheorem#1#2{\trivlist\item[\hskip\labelsep{\bf #1\ #2}]}
\makeatother
\makeatletter %

\newlength{\filength}
\newsavebox{\gcbox}
\sbox{\gcbox}{\framebox[\filength]{\rule{0ex}{2ex}}}

\newlength{\leftjustindent}
\newlength{\@leftjustindent}
\def\leftjust{\let\\\@leftjustcr\let\end\@endleftjust
  \addtolength{\@leftjustindent}{\leftjustindent}
  \vcenter\bgroup
  \halign\bgroup
    \hbox to\displaywidth{
      \rule{\@leftjustindent}{0ex}$\displaystyle##$\hfill
      }\crcr
}
\def\endleftjust{\crcr\egroup\egroup\endgroup}
\def\@endleftjust#1{\crcr\egroup\egroup\@checkend{#1}\endgroup}
\def\@leftjustcr{\crcr}

\newtheorem{theorem}{Theorem}[section]

\newcommand{\qedblob}{\mbox{\rule[-1.5pt]{5pt}{10.5pt}}}
\def\literalqed{{\ \nolinebreak\hfill\mbox{\qedblob\quad}}}

\def\qed{\literalqed}

\newtheorem{lemma}[theorem]{Lemma}

\newcommand{\singlespacing}{\let\CS=
\@currsize\renewcommand{\baselinestretch}{1}\tiny\CS}
\newcommand{\singlespacingplus}{\let\CS=
\@currsize\renewcommand{\baselinestretch}{1.25}\tiny\CS}
\newcommand{\doublespacing}{\let\CS=
\@currsize\renewcommand{\baselinestretch}{1.75}\tiny\CS}
\newcommand{\draftspacing}{\let\CS=
\@currsize\renewcommand{\baselinestretch}{1.65}\tiny\CS}
\makeatother%

\newtheorem{definition}[theorem]{Definition}

\makeatletter
\clubpenalty=\@highpenalty
\widowpenalty=\@highpenalty
\makeatother

\makeatletter
\newcommand{\niceonespacing}{\let\CS=\@currsize\renewcommand{\baselinestretch}{1.1}\tiny\CS}\newcommand{\nicetwospacing}{\let\CS=\@currsize\renewcommand{\baselinestretch}{1.2}\tiny\CS}
\newcommand{\nicethreespacing}{\let\CS=\@currsize\renewcommand{\baselinestretch}{1.3}\tiny\CS}
\newcommand{\singlespacingplusplus}{\let\CS=\@currsize\renewcommand{\baselinestretch}{1.35}\tiny\CS}
\newcommand{\nicefivespacing}{\let\CS=\@currsize\renewcommand{\baselinestretch}{1.5}\tiny\CS}
\newcommand{\nicesixspacing}{\let\CS=\@currsize\renewcommand{\baselinestretch}{1.6}\tiny\CS}
\newcommand{\nicefoospacing}{\let\CS=\@currsize\renewcommand{\baselinestretch}{1.7}\tiny\CS}
\newcommand{\normalspacing}{\niceonespacing}
\makeatother

\normalspacing

\makeatletter
\def\@cite#1#2{[#1\if@tempswa , #2\fi]}
\makeatother

\makeatletter
\def\@citex[#1]#2{\if@filesw\immediate\write\@auxout{\string\citation{#2}}\fi
  \def\@citea{}\@cite{\@for\@citeb:=#2\do
    {\@citea\def\@citea{,\linebreak[0]}\@ifundefined
       {b@\@citeb}{{\bf ?}\@warning
       {Citation `\@citeb' on page \thepage \space undefined}}%
\hbox{\csname b@\@citeb\endcsname}}}{#1}}
\makeatother

\newcommand{\up}{{\rm UP}}

\newcommand{\p}{{\rm P}}

\newcommand{\bh}{{\rm BH}}

\newcommand{\np}{{\rm NP}}

\newcommand{\pij}{{\p^{\bh_i:\bh_j}}}
\newcommand{\pji}{{\p^{\bh_j:\bh_i}}}

\def\pair#1{{{\langle\!\!~#1~\!\!\rangle}}}

\newcommand{\sigmastar}{\mbox{$\Sigma^\ast$}}

\newcommand{\condition}{\,\nottoobig{|}\:}

\def\nats{\naturalnumber}
\newcommand{\naturalnumber}{\ensuremath{{  \mathbb{N} }}}
\def\wit#1{{\mbox{\rm{}WIT}_M(#1)}}

\begin{document}

\bibliographystyle{alpha}

\title{One-Way Functions in Worst-Case Cryptography:\\
Algebraic and Security Properties\thanks{
Supported in part 
by grant
NSF-INT-9815095/\protect\linebreak[0]DAAD-315-PPP-g{\"u}-ab.
Written in part while the second author was
visiting Friedrich-Schiller-Universit\"at Jena and while
the fourth author was
visiting the University of Rochester and the 
Rochester Institute of Technology.} }

\author{
Alina Beygelzimer,  Lane A. Hemaspaandra,  Christopher M. Homan  \\
Department of Computer Science \\
University of Rochester \\
Rochester, NY 14627 
\and
J\"{o}rg Rothe \\
Institut f\"ur Informatik \\
Friedrich-Schiller-Universit\"at Jena\\
07740 Jena, Germany
}

\date{}

\maketitle

\begin{abstract}
We survey recent developments in the study of
(worst-case) one-way functions having strong
algebraic and security properties.
According to~\cite{rab-she:tSPECIALJCSS:aowf}, this line of research
was initiated
in 1984 by Rivest and Sherman who designed two-party
secret-key agreement protocols that use strongly noninvertible, total,
associative one-way functions as their key building blocks.  If
commutativity is added as an ingredient, these protocols can be used
by more than two parties, as noted by Rabi and
Sherman~\cite{rab-she:tSPECIALJCSS:aowf} who also developed
digital signature protocols that are based on such enhanced
one-way functions.

Until recently, it was an open question whether one-way functions
having the algebraic and security properties that these protocols
require could be created from any given one-way function.  Recently,
Hemaspaandra and Rothe~\cite{hem-rot:j:aowf} resolved this open issue
in the affirmative,
by showing that one-way functions exist if and only if 
strong, total, commutative, associative one-way functions exist.

We discuss this result, and the work of Rabi, Rivest, and Sherman, and 
recent work of Homan~\cite{hom:m:aowf} that makes progress on related
issues.  

\end{abstract}

\section{Motivation}
\label{s:motiv}

\begin{description}
\item[Professor One:]
Hello, Professor Way! How's life?

\item[Professor Way:]
Very exciting indeed. I've developed some very exciting worst-case 
cryptographic protocols. If you read these papers and manuscripts of
mine, you'll see how intuitively attractive, interesting, and 
exciting my protocols are.

\item[Professor One] (spends 10 minutes skimming the papers as 
Professor Way waits patiently){\bf :}
Wow... I {\em am\/} attracted, interested, and excited by those
protocols. But wait. Is there some catch?

\item[Professor Way:]
Well, I do assume that we have, to use in the protocols, (worst-case)
one-way functions that have various additional algebraic and
security properties such as associativity, commutativity, and
``strong'' noninvertibility.

\item[Professor One:]
You're assuming {\em WHAT\/}!!?? Whether vanilla one-way functions
exist is a major open research issue, and you're throwing in all
sorts of wild extra requirements on one-way functions?
Though like many people I believe that vanilla one-way functions
exist, I have no similar intuition as to whether one-way functions
exist with the many extra properties you are assuming. And so, I must
view your protocols as less attractive than protocols built on the
assumption that vanilla one-way functions exist.
\end{description}

\bigskip

\noindent
(Until recently, Professor Way would not have had any good reply
at this point. However, due to the work this article is about,
Professor Way does have a slam-dunk reply.)

\bigskip

\begin{description}
\item[Professor Way:]
Your worries are completely natural, but nonetheless unfounded.
The reason is that one can now prove that all those ``wild''
extra properties come for free. That is, it remains an open issue
whether vanilla one-way functions exist. And it also remains an 
open issue whether spiffy (say, strongly noninvertible, total,
commutative, associative) one-way functions exist. However, they
are the same open issue: {\em Spiffy one-way functions exist if
and only if vanilla one-way functions exist}.
\end{description}

\section{Organization and Definitions}
\label{s:definitions}

Section~\ref{s:motiv} provided an example of why it may be
useful to understand the interactions between one-way-ness and other
properties. The present section gives the basic formal
definitions. Section~\ref{s:progress} summarizes the main results of
the papers we survey. Section~\ref{s:proofs} sketches proofs of
restricted cases of some of the results discussed.

We now define the concepts important to this
survey. 

Throughout this paper, we mainly deal with 2-ary functions, 
in particular functions mapping
from $\sigmastar \times \sigmastar$ to~$\sigmastar$, where $\Sigma =
\{0,1\}$ is our fixed alphabet.  We use both prefix and infix notation
for 2-ary functions~$\sigma$, i.e., $\sigma(x,y) = x \sigma y$.
Unless explicitly stated as being total or one-to-one, the functions
we consider are partial and potentially many-to-one.  
We
assume that we have a
pairing
function $\pair{\cdot , \cdot}$ mapping 
$\sigmastar \times \sigmastar$ onto~$\sigmastar$ with the standard
nice properties.

Worst-case one-way functions have been studied by many researchers, see,
e.g., the
papers~\cite{gro-sel:j:complexity-measures,ko:j:operators,sel:j:one-way,rab-she:j:aowf,hem-rot:j:aowf,hom:m:aowf}. Definition~\ref{def:one-way}
presents the case of 2-ary one-way functions.

\begin{definition} 
\label{def:one-way}
(see, e.g., \cite{rab-she:j:aowf})~
For any 2-ary function 
$\sigma : \sigmastar \times \sigmastar \rightarrow \sigmastar$, 
we say:
\begin{itemize}
\item $\sigma$ is {\em honest\/} if $\sigma$ does not shrink 
its inputs more than
by a polynomial amount, i.e., there is a polynomial $p$ such that for 
every image element $c$ of~$\sigma$, there is a domain element 
$(a,b)$ of $\sigma$ satisfying $a \sigma b = c$ and $|a| + |b| \leq p(|c|)$;

\item $\sigma$ is
({\em polynomial-time\/}) {\em invertible\/} if there exists a 
total, polynomial-time computable function \mbox{$g : \sigmastar
\rightarrow \sigmastar \times \sigmastar$} such that for every
$c$ in the image of~$\sigma$, $\sigma(g(c)) = c$;

\item $\sigma$ is a {\em one-way function\/} if $\sigma$ is honest, 
polynomial-time computable, and noninvertible. 
\end{itemize}
\end{definition}

As we will see in Section~\ref{s:progress}, if one-way functions
possess certain algebraic properties such as associativity and
commutativity, they may be useful as building blocks of some clever
cryptographic protocols designed by Rivest, Rabi, and
Sherman. The following definition is due to Hemaspaandra and
Rothe~\cite{hem-rot:j:aowf}.\footnote{Rabi and
Sherman~\cite{rab-she:j:aowf} use a different notion dubbed ``weak
associativity'' in~\cite{hem-rot:j:aowf}: Any 2-ary function $\sigma$
is said to be {\em weakly associative\/} if the equality $a \sigma (b
\sigma c) = (a \sigma b) \sigma c$ holds for all $a,b,c \in
\sigmastar$ {\em satisfying that both $(a,b)$ and $(b,c)$ are in the
domain of $\sigma$ and if $(a,b)$ and $(b,c)$ are in the domain of
$\sigma$ then so are $(a,b \sigma c)$ and $(a \sigma b, c)$}. (Rabi
and Sherman actually quantify over all $a,b,c \in \sigmastar$ {\em
satisfying that each of $(a,b)$, $(b,c)$, $(a,b \sigma c)$, and $(a
\sigma b, c)$ is in the domain of~$\sigma$}, a phrasing that is
logically equivalent with our phrasing, but that may contain terms
that are not well-defined: $(a \sigma b, c)$ is not well-defined if
$\sigma$ is not defined at $(a,b)$.)~~The distinction between these
two notions of associativity, in brief, can be explained via
Kleene's~\cite[pp.~327--328]{kle:b:metamathematics} distinction
between complete equality and weak equality for partial functions; 
see~\cite{hem-rot:j:aowf} for a discussion of some
weaknesses of weak associativity.  
\label{foo:weak-ass}
}

\begin{definition} 
\label{def:associative}
For any 2-ary function 
$\sigma : \sigmastar \times \sigmastar \rightarrow \sigmastar$,
define the set $\Gamma = \sigmastar \cup \{\bot\}$ 
and an extension $\widehat{\sigma} : \Gamma \times
\Gamma \rightarrow \Gamma$
of $\sigma$ as follows:\footnote{A 
change made by a journal copyeditor inserted a typo into 
Definition~2.3 of~\cite{hem-rot:j:aowf}.
Line~27 of page~651 of~\cite{hem-rot:j:aowf} should correctly read
as equation~(\ref{eq:typo}) given here (note the 
occurrence of ``$a \neq \bot$'' rather than the typo ``$a\not=1$'').
}
\begin{equation}
  \widehat{\sigma}(a,b) = \left\{ 
\begin{array}{ll} \sigma(a,b)  &
  \mbox{if $a \neq \bot$ and $b \neq \bot$ and $(a,b) \in \mbox{\rm
  domain}(\sigma)$} \\ 
  \bot & \mbox{otherwise.}
\end{array} 
\right.
\label{eq:typo}
\end{equation}
We say $\sigma$ is {\em associative\/} if 
$(a \widehat{\sigma} b) \widehat{\sigma} c = 
a \widehat{\sigma} (b \widehat{\sigma} c)$ holds for all
$a,b,c \in \sigmastar$.
We say $\sigma$ is {\em commutative\/} if 
$a \widehat{\sigma} b = b \widehat{\sigma} a$ holds 
for all $a,b \in \sigmastar$.
\end{definition}

Rabi and Sherman~\cite{rab-she:j:aowf} use a notion of strong
noninvertibility: A 2-ary function $\sigma$ is strongly noninvertible
if even given the output and an argument, computing the other argument 
is not a polynomial-time achievable task.

Let us state this formally (see~\cite{rab-she:j:aowf,hem-rot:j:aowf}).

\begin{definition} 
\label{def:strong}
A 2-ary function 
$\sigma : \sigmastar \times \sigmastar \rightarrow \sigmastar$ 
is said to be {\em strong\/} if no polynomial-time
computable function $g : \sigmastar
\rightarrow \sigmastar$ satisfies either of the following two conditions:
\begin{itemize}
\item For all $c$ in the image of $\sigma$ and for all $a \in \sigmastar$,
if there is some $b \in \sigmastar$ with $a \sigma b = c$, then
$\sigma(a,g(\pair{a,c})) = c$.

\item For all $c$ in the image of $\sigma$ and for all $b \in \sigmastar$,
if there is some $a \in \sigmastar$ with $a \sigma b = c$, then
$\sigma(g(\pair{b,c}),b) = c$.
\end{itemize}
\end{definition}

Note that strongness implies noninvertibility.

Finally, we define bounded ``many-to-one''-ness.
Denote the set of nonnegative integers by~$\nats$. 

\begin{definition} 
\label{def:m-to-1}
Let $h : \nats \rightarrow \nats$ be any total function
and let $\sigma : \sigmastar \times \sigmastar \rightarrow \sigmastar$
be any function.
We say $\sigma$ is {\em $h(k)$-to-one\/} if for every $b$ of length
$k$ in the image of~$\sigma$, the cardinality\footnote{Throughout 
this paper, for any function $f$
(even if $f$ happens to be one-to-one)
and for any image element $z$ of~$f$, we mean by ``the preimage 
of $z$ under~$f$'' the set of all domain elements
mapped to $z$ by~$\sigma$.
}
of the preimage of $b$ under $\sigma$ is at most~$h(k)$.
\end{definition}

\section{Progress on Algebraic and Security Properties for 
One-Way Functions in Worst-Case Cryptography}
\label{s:progress}

\subsection{Rabi and Sherman: Weakly Associative One-Way Functions Exist 
If and Only If One-Way Functions Exist}
\label{s:progress-rab-she}

The ``original'' result about one-way functions 
is:

\begin{theorem}
\label{thm:folk}
{\rm{}(see~\cite{bal-dia-gab:b:sctI:95} 
and~\cite[Proposition~1]{sel:j:one-way})}\quad
$\p \neq \np$ if and only if one-way functions 
exist.\footnote{This 
result is widely known and cited, but the authors have yet to find 
an attribution as to who first discovered it.

For the special case of one-to-one one-way functions
(see the excellent survey 
by Selman~\cite{sel:j:one-way}), the history
is much clearer.
The analogous theorem for those is the following.
\protect\begin{theorem}
\label{thm:folk-up}
{\rm{}\cite{gro-sel:j:complexity-measures,ko:j:operators,ber:thesis:iso}}\quad
One-to-one one-way functions exist if and only if $\p \neq \up$, where
$\up$ is Valiant's unambiguous polynomial time~\cite{val:j:checking}.
\end{theorem}
This theorem was found independently by 
Grollmann and Selman~\cite{gro-sel:j:complexity-measures} and
Ko~\cite{ko:j:operators}, and 
Berman's 
thesis~\cite{ber:thesis:iso} independently obtained essentially the 
same result (see~\cite{sel:j:one-way}). 

To avoid possible confusion, we mention that though our
Definition~\ref{def:one-way} (and this entire article) does not
require one-way functions to be one-to-one, some authors do
mean ``one-to-one one-way function'' when they write 
``one-way function.''  
\label{foo:folk-up}
} %
\end{theorem}

However, writers of (even worst-case) cryptographic protocols began to
desire stronger building blocks than these vanilla one-way
functions---in particular, one-way functions with enhanced algebraic
and security properties. In fact, according
to~\cite{rab-she:tSPECIALJCSS:aowf}, this idea was suggested in 1984
by Rivest and Sherman with respect to secret-key agreement.

This excellent, insightful idea of Rivest and Sherman led to the 
important 1993 paper of Rabi
and Sherman (\cite{rab-she:tSPECIALJCSS:aowf}, see also the journal
version~\cite{rab-she:j:aowf}), which proposes explicit protocols that
exploit such algebraic and security properties as strong
noninvertibility, totality, commutativity, and weak associativity.
This of course raised the issue of whether one-way functions with
these properties were likely to exist. Rabi and Sherman prove the
following result.

\begin{theorem}
\label{thm:rab-she-1}
{\rm{}\cite{rab-she:tSPECIALJCSS:aowf,rab-she:j:aowf}}\quad
Weakly associative, commutative one-way functions exist if and only if 
one-way functions exist.
\end{theorem}

Interestingly, their proof technique is quite different from the
techniques used to study one-to-one one-way functions.  

Note, however,
that the proof of Theorem~\ref{thm:rab-she-1} does not achieve
totality, 
associativity
(as per Definition~\ref{def:associative}),
or strongness.
Another result due to Rabi and Sherman is the following.

\begin{theorem}
\label{thm:rab-she-2}
{\rm{}\cite{rab-she:tSPECIALJCSS:aowf,rab-she:j:aowf}}\quad
No total, one-to-one, weakly associative one-way functions exist.
\end{theorem}

\subsection{Hemaspaandra and Rothe: Strong, Total, Commutative, Associative
One-Way Functions Exist If and Only If One-Way Functions Exist}
\label{s:progress-hem-rot}

One key worry with the protocols discussed by Rabi and Sherman is that
their key characterization result, Theorem~\ref{thm:rab-she-1}, is not
strong enough to ensure that (with at least the same certainty as that
with which vanilla one-way functions exist) there exist one-way
functions having the properties the protocols of Rabi, Rivest, and
Sherman require. For example,
strong noninvertibility is important for the protocols, and a lack of totality
would severely decrease their applicability.

Hemaspaandra and Rothe remove this worry by proving that spiffy
one-way functions are just as likely to exist as vanilla one-way 
functions.  In particular, they prove the following result. 

\begin{theorem}
\label{thm:hem-rot-1}
{\rm{}\cite{hem-rot:j:aowf}}\quad
Strong, total, commutative, associative one-way functions exist
if and only if one-way functions exist.
\end{theorem}

\smallskip

\begin{description}
\item[Professor One:]
Gotcha!  Theorem~\ref{thm:hem-rot-1} is about {\em associative\/}
one-way functions as in Definition~\ref{def:associative}, yet the
protocols of Rivest et al.\ require {\em weakly associative\/}
one-way functions.  And in one of your overlong footnotes you claim
that weak associativity is different than associativity by which,
I suppose, you mean {\em provably\/} different.

\item[Professor Way:]
That's right. But, firstly, every associative function
outright is weakly associative, so Theorem~\ref{thm:hem-rot-1}
{\em does\/} provide the type of one-way function
needed for the protocols.  Secondly, for {\em total\/} 
2-ary functions such as those of Theorem~\ref{thm:hem-rot-1},
the two notions of associativity coincide anyway;
look at~\cite[Proposition~2.4]{hem-rot:j:aowf} if you don't see 
why these claims hold.
Thirdly, note that most results of~\cite{hem-rot:j:aowf} 
and of~\cite{rab-she:j:aowf} are shown, in~\cite{hem-rot:j:aowf}, 
to hold both for associative and weakly associative one-way
functions. And finally: Motivation time is over, we are in the 
middle of a technical section, so the two of us shouldn't distract
the reader from reading the results and proof sketches.
\end{description}

\bigskip

The proof of Theorem~\ref{thm:hem-rot-1}, which will be partially
discussed in Section~\ref{s:proofs}, has two parts. One part shows how
to establish strongness, associativity, and commutativity.  The second
part shows how the very special strong, associative, and commutative
one-way function created from any given one-way function in the first
part of the proof can be extended to achieve totality without
destroying any of the other properties.

Note that this extension is a very specific ``conversion to
totality.''  Another result 
of~\cite{hem-rot:j:aowf}
addresses the issue of broader
``conversions to totality.''  In particular, \cite{rab-she:j:aowf}
gives a construction, call it~$C$, that it asserts lifts any nontotal,
weakly associative one-way function whose domain is in P to a total,
weakly associative one-way function. Though it remains possible that
this construction in fact always works, under a plausible
complexity-theoretic hypothesis Hemaspaandra and Rothe~\cite{hem-rot:j:aowf}
show that there will be cases on which it fails.

\begin{theorem}
\label{thm:hem-rot-2}
{\rm{}\cite{hem-rot:j:aowf}}\quad
If $\up \neq \np$ then there exists a weakly associative one-way function
$\tau$ such that
\begin{description}
\item[(a)] the domain of $\tau$ is in~$\p$,

\item[(b)] there exists some $x \in \sigmastar$ such that $(x,x)$ is 
not in the domain of~$\tau$, and

\item[(c)] construction~$C$ fails on~$\tau$, that is, 
the total extension of $\tau$ yielded by $C$
is not weakly associative.
\end{description}
\end{theorem}

Note that, for construction~$C$ to work, both condition~(a) and
condition~(b) are required.  While in~\cite{rab-she:j:aowf}, without
proof, condition~(b) is simply assumed to be true for {\em every\/}
nontotal, weakly associative one-way function, there may well be
counterexamples to this claim.  However, for the particular function
$\tau$ constructed in the proof of Theorem~\ref{thm:hem-rot-2},
condition~(b) is explicitly shown to hold. Thus, construction~$C$ does
not fail on $\tau$ (see condition~(c)) because it cannot be applied
to~$\tau$, but rather because $C$ does not preserve weak
associativity. In contrast, $C$ does preserve associativity as defined
in Definition~\ref{def:associative} and so is useful in achieving the
``conversion to totality'' in the second part of the proof of
Theorem~\ref{thm:hem-rot-1}.

Finally, what about the issue of
injectivity (i.e., one-to-one-ness) for associative one-way functions?
Theorem~\ref{thm:rab-she-2}, due to Rabi and Sherman, states
that no total, weakly associative function (and so, by the above
comment of Professor Way, no total, associative function) is
injective.  However, if one does not require totality then associative,
injective one-way functions are no less likely to exist than injective
one-way functions, which expands Theorem~\ref{thm:folk-up}.

\begin{theorem}
\label{thm:hem-rot-3}
{\rm{}\cite{hem-rot:j:aowf}}\quad
One-to-one, associative one-way functions exist
if and only if one-to-one one-way functions exist.
\end{theorem}

Hemaspaandra and Rothe~\cite{hem-rot:j:aowf}
also establish that equivalent to the two conditions of
Theorem~\ref{thm:hem-rot-3} (and thus to the condition ``$\p \neq \up$,''
see Theorem~\ref{thm:folk-up}) is the existence of strong, commutative,
associative one-way functions that satisfy a certain weak notion of
injectivity called ``unordered injectivity.''
\begin{definition}
A 2-ary function is
{\em unordered-injective\/} if for all $a,b,c,d \in \sigmastar$ with
$(a,b)$ and $(c,d)$ in the domain of~$\sigma$, $\sigma(a,b) =
\sigma(c,d)$ implies $\{a,b\} = \{c,d\}$.
\end{definition}
They left open the issue of
whether for total, associative functions---which cannot be one-to-one
by Theorem~\ref{thm:rab-she-2}---also two-to-one-ness is precluded,
and what bounds on the ``many-to-one''-ness of such functions (one-way
or otherwise) can be shown to hold.  The next section gives an answer
to the first question and reports on recent progress towards
resolving the general case.

\subsection{Homan: Amount of ``Many-to-One''-ness and its Interaction
with Algebraic and Security Properties}
\label{s:progress-hom}

Suppose we can encode a message using an associative one-way function,
and its intended recipient can decode it.  Can the space to which the
encrypted message is mapped by the decoding function be feasibly
searched---or is it a haystack?  What if the number of potential
decodings of the encoded message is so large that it cannot be
determined in polynomial time which decoding was the original message?
As mentioned in Footnote~\ref{foo:folk-up}, some researchers require
one-way functions to always be one-to-one. Others merely require that
the ambiguity of the possible decodings be polynomially bounded, so
that they can be efficiently searched. In particular, Allender and
Rubinstein~\cite{all-rub:j:print,all:coutdatedExceptForPUNCstuff:complexity-sparse}
introduce ``poly-to-one'' one-way functions and prove an analog of
Theorem~\ref{thm:folk} for those functions
(see also~\cite{rot-hem:tCORR:one-way} for an expansion of their result),
and 
Watanabe~\cite{wat:j:hardness-one-way},
Hemaspaandra and Hemaspaandra~\cite{hem-hem:j:quasi}, and others
have
studied variations of constant-bounded ambiguity.
But how does bounded ``many-to-one''-ness, or even one-to-one-ness,
interact with algebraic and security properties such as associativity
and strongness?

We have already seen that---whether or not one-way-ness is
involved---associativity and totality preclude one-to-one-ness
(Theorem~\ref{thm:rab-she-2}).  Homan~\cite{hom:m:aowf} strengthens
this result.

\begin{theorem}
\label{thm:hom-1}
{\rm{}\cite{hom:m:aowf}}\quad
No total, associative function is constant-to-one.
\end{theorem}

Homan also proves that this bound is tight by providing the following
upper bound: For each nondecreasing, unbounded function~$g$, there
exists an $\mathcal{O}(g)$-to-one, total, commutative, associative
function.

Now, let us throw one-way-ness in and ask again: What bounds can one
prove on the ``many-to-one''-ness of one-way functions having the
algebraic and security properties surveyed in this article? 

\begin{theorem}
\label{thm:hom-2}
{\rm{}\cite{hom:m:aowf}}\quad
If $\p \neq \up$ then there exists an
$\mathcal{O}(n)$-to-one, strong, total, associative one-way function.
\end{theorem}

Regarding lower bounds, Homan establishes 
the following result. 

\begin{theorem}
\label{thm:hom-3}
{\rm{}\cite{hom:m:aowf}}\quad
For every total, honest, associative function $\sigma$ 
whose output length is bounded by a polynomial in the length of the input, 
there exists an $m \in \nats$ such that $\sigma$ is not 
$o(f^{-1})$-to-one, where 
$f(x) = \lceil 2 \log x \rceil^{m^{\lceil \log x \rceil}}$.
\end{theorem}

There is a rather wide gap between this lower bound and the 
upper bound given
in Theorem~\ref{thm:hom-2} (under a plausible complexity-theoretic
hypothesis).  That is, there is a gap between the
slowest \textit{known} growth-rate of the ``many-to-one''-ness of
strong, total, associative one-way functions and their slowest
\textit{possible} growth-rate.  Closing this gap is an interesting
open issue. Also open is the degree of  
``many-to-one''-ness for \textit{commutative}, strong, total,
associative one-way functions.

\section{Proof Sketches}
\label{s:proofs}

In this section, we present proof sketches for some of the results 
surveyed and give the flavor of some of the different techniques used.

\subsection{Proof Sketches Related to Hemaspaandra and Rothe's Work}
\label{s:proofs-hem-rot}

\noindent
{\bf Proof Sketch of Theorem~\ref{thm:hem-rot-1}.}~~Since every spiffy
one-way function is a very particular vanilla one-way function, it is
enough to show how to create, given any vanilla one-way function~$v$,
a one-way function that is strong, total, commutative, and
associative.  By Theorem~\ref{thm:folk}, we can just as well create this
function from the assumption that $\p \neq \np$.  (See Grollmann and
Selman~\cite{gro-sel:j:complexity-measures} for how to convert any
given one-way function into a set in $\np$ that is not in~$\p$.
Although this conversion in Grollmann and Selman is done
for 1-ary one-to-one one-way functions and~$\p$ versus~$\up$, the analogous
approach works cleanly for the case of 2-ary many-to-one one-way functions
and~$\p$ versus~$\np$.)

So, given~$v$, let $A_v$ be the corresponding set in $\np - \p$. 
We will now define the little
brother---call him~$\sigma$---of the spiffy one-way function we are
going to construct from~$A_v$.  Think of $\sigma$ as a piece of
Swiss cheese, full of plenty of
delicious, tasty, carefully made cheese, but also full of
holes. That is, $\sigma$ will be a strong, commutative, associative
one-way function, but it will in fact not be total.  The big brother of
$\sigma$, then, will be the same piece of Swiss cheese, still
delicious, tasty, and carefully made, but with its holes plugged. That
is, it will be the total extension of $\sigma$---carefully preserving
each of $\sigma$'s algebraic and security properties---that is yielded
by construction~$C$ mentioned in Section~\ref{s:progress-hem-rot}.  In
this survey, we restrict ourselves to making just the Swiss cheese
$\sigma$ with holes.

How do we make~$\sigma$? First, forget about $\sigma$ being a piece of Swiss
cheese. Rather, imagine $\sigma$ to be a police officer at work.  

It is a busy morning at the police department.
Officer~$\sigma$ has many reports on her desk describing incidents $x$
that happened last night. Our set $A_v \in \np - \p$ will be the set
of all incidents that are crimes. (Suppose that, 
every night, many crimes happen
and most of them are rather difficult to solve.)~~A report on
Officer~$\sigma$'s desk may contain the description of an incident~$x$
with a file copy attached to it (such a report has the form
$\pair{x,x}$).  Another report may contain the description of a
crime~$x$ with an eye witness's statement $w$ attached to it
(such a report has the form $\pair{x,w}$).  There are all sorts of
other reports as well.

Luckily, Officer~$\sigma$ can easily tell incident descriptions
apart from witness statements, so she always knows whether the report
at hand is of the form $\pair{x,x}$ or $\pair{x,w}$.
Also, Officer~$\sigma$ can easily check how reliable a 
witness is, since they use lie detectors at this police department
to verify each witness statement taken.

Every once in a while, Officer~$\sigma$ grabs two reports 
$a$ and~$b$ (one with her left hand and one with her right hand), reads 
them both, and chooses one of $a$ and $b$ to pass
on to her boss, Sgt.~{\protect $\hat{\sigma}$}, 
dumping the other one. Sometimes, she dumps them both.
Here is how Officer~$\sigma$ makes her decision on which reports to pass
on and which to dump:
\begin{itemize}
\item Whenever report $a$ is of the form $\pair{x,w_1}$ and report $b$
is of the form $\pair{x,w_2}$ (that is, both describe the same
incident~$x$, which appears to be a crime, for there are
two---possibly identical---witness statements attached to it),
Officer~$\sigma$ picks one of $a$ and $b$ to pass on to
Sgt.~{\protect $\hat{\sigma}$}, dumping the other one.  In
particular, she always passes on the report containing the {\em
shorter\/} (to be more precise, the lexicographically lesser) witness 
statement.

\item Whenever one of the reports has the form $\pair{x,x}$ and the
other one has the form $\pair{x,w}$ for the same crime~$x$, where $w$
is a witness statement for~$x$, Officer~$\sigma$ passes report
$\pair{x,x}$ on to Sgt.~{\protect $\hat{\sigma}$}, distractedly
dumping $\pair{x,w}$ into the waste
basket.\footnote{That in part explains why so
few crimes are solved in this town.}

\item Whenever the reports are not of the form described in the above
two cases, Officer~$\sigma$ dumps them both.
\end{itemize}

\bigskip

Now, let us be a bit more formal. 
A witness for ``$x \in A_v$'' is any string
$w \in \sigmastar$ encoding an accepting path of $M$ on input~$x$,
where $M$ is a fixed NP machine accepting~$A_v$.
For each $x \in A_v$, define the set of witnesses for ``$x \in A_v$'' 
by
\[
\wit{x} = \{ w \in \sigmastar \condition \mbox{$w$ is a witness
for ``$x \in A_v$''} \}.
\]

We may assume that, for each $x \in A_v$, any witness $w$ for ``$x \in
A_v$'' is of length $p(|x|)$ for some strictly increasing
polynomial~$p$, and the length of $w$ is strictly larger than the
length of~$x$. This assumption is just a technical detail that enables
Officer $\sigma$ to tell input strings in $A_v$ apart from their witnesses, 
a property that will be useful later on.

Given any two strings $a$ and $b$ in~$\sigmastar$,
define $\sigma(a,b)$ as follows:
\begin{itemize}
\item If there is some
$x \in \sigmastar$ for which there exist witnesses $w_1,w_2 \in \wit{x}$
such that $a = \pair{x,w_1}$ and $b = \pair{x,w_2}$, then $\sigma(a,b)$
is defined to be the string $\pair{x,\min(w_1,w_2)}$, 
where $\min(w_1,w_2)$ denotes the lexicographically smaller of $w_1$
and~$w_2$.

\item If there is some
$x \in \sigmastar$ for which there exists some witness $w \in \wit{x}$
such that $a = \pair{x,x}$ and $b = \pair{x,w}$, or $a = \pair{x,w}$
and $b = \pair{x,x}$, then $\sigma(a,b)$
is defined to be the string $\pair{x,x}$.

\item Otherwise, $\sigma(a,b)$ is undefined, that is, 
there is a hole in the domain of $\sigma$ at~$(a,b)$.
\end{itemize}

It remains to prove that $\sigma$ has the desired properties.  That
$\sigma$ is honest and commutative is immediate.  That $\sigma$ is
polynomial-time computable can be seen as follows.  By our assumption
that for each $x$ in~$A_v$, the length of any witness string for ``$x
\in A_v$'' is strictly larger than the length of~$x$, there is no
ambiguity in deciding whether $\sigma$'s arguments, $a$ and~$b$, are
of the form $\pair{x,x}$ or $\pair{x,w}$, where $w$ is a potential
witness for ``$x \in A_v$.''~~Moreover, we can of course decide
in polynomial time whether a potential witness $w$ for ``$x \in A_v$''
indeed is a witness.

The strongness of $\sigma$ is shown by way of contradiction.  Suppose
there is a polynomial-time computable function $g$ such that, for any
string $c$ in the image of $\sigma$ and for any fixed first argument
$a \in \sigmastar$ for which there is some second argument $b \in
\sigmastar$ with $a \sigma b = c$, it holds that
$\sigma(a,g(\pair{a,c})) = c$.  Using~$g$, one could then decide $A_v$
in polynomial time as follows:
\begin{quote}
Given any input string~$x$, to decide whether or not $x$ is in~$A_v$,
compute the string $g(\pair{\pair{x,x},\pair{x,x}})$. Compute the
projections, say $u$ and $w$, of our pairing function at
$g(\pair{\pair{x,x},\pair{x,x}})$; that is, compute the unique strings
$u$ and $w$ for which $\pair{u,w} =
g(\pair{\pair{x,x},\pair{x,x}})$. Accept $x$ if and only if $u = x$
and $w \in \wit{x}$.
\end{quote}
This polynomial-time algorithm for $A_v$ contradicts our assumption
that $A_v$ is not in~$\p$.  Hence, $\sigma$ cannot be inverted in
polynomial time even if the first argument is given.  An analogous
argument shows that no polynomial-time computable function can invert
$\sigma$ even if the second argument is given.  Hence, $\sigma$ is
strong.

It remains to show that $\sigma$ is associative.
Let $a, b, c \in \sigmastar$ be any fixed arguments for~$\sigma$.
Let the projections of our pairing function at
$a$, $b$, and $c$ be given by
$a = \pair{a_1, a_2}$, $b = \pair{b_1, b_2}$, and
$c = \pair{c_1, c_2}$.
Let $k \in \{0, 1, 2, 3\}$ be the number that tells you 
how many of $a_2$, $b_2$, 
and~$c_2$ are elements of $\wit{a_1}$. For example, if $a_2 = c_2
\in \wit{a_1}$, but $b_2 \not\in \wit{a_1}$, then $k = 2$.

According to Definition~\ref{def:associative}, we have to show
that 
\begin{eqnarray}
\label{equ:ass}
(a \widehat{\sigma} b) \widehat{\sigma} c = a \widehat{\sigma} (b
\widehat{\sigma} c) ,
\end{eqnarray}
where $\widehat{\sigma}$ is the extension of $\sigma$ from that
definition.

There are two cases.

\begin{description}
\item[Case~1:] Suppose $a_1 = b_1 = c_1$ and $\{a_2, b_2, c_2 \}
\subseteq \{a_1\} \cup \wit{a_1}$.  The intuition in this case is that
$\sigma$ decreases by one the number of witnesses that may occur in
its arguments in the following way.

If zero of $\sigma$'s arguments contain a witness for ``$a_1\in A$,''
then $\sigma$ is undefined, so $\widehat{\sigma}$ outputs~$\bot$.

If exactly one of $\sigma$'s arguments contains a witness for
``$a_1\in A$,'' then $\sigma$---and thus $\widehat{\sigma}$ as
well---has the value $\pair{a_1, a_1}$.

If both of $\sigma$'s arguments contain a witness for ``$a_1\in A$,''
then $\widehat{\sigma}$ outputs $\pair{a_1, w}$, where $w \in \{a_2,
b_2, c_2\}$ is the lexicographically smaller of the two witnesses.

From the above we may conclude the following.

If $k \in \{0,1\}$ then
$(a \widehat{\sigma} b) \widehat{\sigma} c = \bot = 
a \widehat{\sigma} (b \widehat{\sigma} c)$.

If $k = 2$ then $(a \widehat{\sigma} b) \widehat{\sigma} c =
\pair{a_1, a_1} = a \widehat{\sigma} (b \widehat{\sigma} c)$.

If $k = 3$ then $(a \widehat{\sigma} b) \widehat{\sigma} c =
\pair{a_1, \min(a_2, b_2, c_2)} =
a \widehat{\sigma} (b \widehat{\sigma} c)$, where
$\min(a_2, b_2, c_2)$ denotes the lexicographically smallest of
$a_2$, $b_2$, and~$c_2$.

In each case, equation~(\ref{equ:ass}) is satisfied.

\item[Case~2:] Suppose case~1 does not hold. This implies that either
$a_1 \neq b_1$ or $a_1 \neq c_1$ or $b_1 \neq c_1$, or
it holds that $a_1 = b_1 = c_1$ and 
$\{a_2, b_2, c_2 \} \not\subseteq \{a_1\} \cup \wit{a_1}$.
In either of these two subcases of case~2, one can verify
that
$(a \widehat{\sigma} b) \widehat{\sigma} c = \bot = 
a \widehat{\sigma} (b \widehat{\sigma} c)$.
Thus, in each subcase, equation~(\ref{equ:ass}) is satisfied.
\end{description}
Hence, $\sigma$ is associative. This completes the proof sketch.~\qed

\medskip

\noindent
{\bf Proof Sketch of Theorem~\ref{thm:hem-rot-2}.}~~Assuming 
$\up \neq \np$, we will show that the ``conversion to totality''
construction of Rabi and Sherman (which was called construction~$C$ 
in Section~\ref{s:progress-hem-rot}) does not preserve weak 
associativity. 

Construction~$C$ works as follows. 
Suppose we are given any nontotal function 
$\tau : \sigmastar \times \sigmastar \rightarrow \sigmastar$
satisfying that (i)~the domain of $\tau$ can be decided in polynomial
time, and (ii)~there exists some string 
$\mbox{\it trashbin} \in \sigmastar$ such that 
$(\mbox{\it trashbin}, \mbox{\it trashbin})$ 
is not in the domain of~$\tau$. Construction~$C$ converts
$\tau$ into a total function 
$\tilde{\tau} : \sigmastar \times \sigmastar \rightarrow \sigmastar$
defined as follows:
\begin{equation}
\tilde{\tau}(a,b) = \left\{ 
\begin{array}{ll} 
\tau(a,b) & \mbox{if $(a,b)$ is in the domain of $\tau$} \\
\mbox{\it trashbin} & \mbox{otherwise,}
\end{array} 
\right.
\label{equ:construction}
\end{equation}
that is, $\mbox{\it trashbin}$ is used to dump 
all garbage elements of $\tau$ (i.e., elements on which
$\tau$ is not defined).

We will now define a 2-ary function $\tau$ that resembles
Officer~$\sigma$ from the proof of
Theorem~\ref{thm:hem-rot-1}. However, unlike~$\sigma$, $\tau$ will be
merely {\em weakly\/} associative. We then show that the total
extension~$\tilde{\tau}$ that is yielded by applying
construction~$C$ to~$\tau$ is not weakly associative.

Pick a set $L$ in $\np - \up$ and a nondeterministic polynomial-time
Turing machine $M$ accepting~$L$.  We assume that all technical
requirements that were useful in defining $\sigma$ also hold in this
proof. In particular, for any $x \in L$, all witnesses for ``$x \in
L$'' are of length greater than the length of~$x$, and $\wit{x}$ is
the set of witnesses for~$x$, defined as in the proof of
Theorem~\ref{thm:hem-rot-1}.

Given any two strings $a$ and $b$ in~$\sigmastar$, define $\tau(a,b)$
as follows:
\begin{itemize}
\item If there is some $x \in \sigmastar$ for which there exists some
witness $w \in \wit{x}$ such that $a = \pair{x,w}$ and $b =
\pair{x,w}$, then $\tau(a,b)$ is defined to be the string
$\pair{x,w}$.

\item If there is some $x \in \sigmastar$ for which there exists some
witness $w \in \wit{x}$ such that $a = \pair{x,x}$ and $b =
\pair{x,w}$, or $a = \pair{x,w}$ and $b = \pair{x,x}$, then
$\tau(a,b)$ is defined to be the string $\pair{x,x}$.

\item Otherwise, $\tau(a,b)$ is undefined, that is, 
there is a hole in the domain of $\tau$ at~$(a,b)$.
\end{itemize}

Note that $\sigma$ and $\tau$ differ only in the first item of their
definitions. It is not difficult to see that $\tau$ is a weakly
associative one-way function. So, it remains to prove that
conditions~(a), (b), and~(c) of Theorem~\ref{thm:hem-rot-2} are
satisfied.

Condition~(a): The domain of $\tau$ can be decided in polynomial time,
since witness checking can be done in deterministic polynomial time
and since we can distinguish between input strings and
their potential witnesses by our length requirement.

Condition~(b): Since $L \not\in \up$, we have $L \neq \sigmastar$; so,
there must be a string $\hat{x}$ not in~$L$.  Let $\mbox{\it trashbin}
= \pair{\hat{x},1\hat{x}}$.  Note that there is no string $x \in
\sigmastar$ for which $\mbox{\it trashbin} = \pair{x,x}$, and there
are no strings $x \in \sigmastar$ and $w \in \wit{x}$ for which
$\mbox{\it trashbin} = \pair{x,w}$ (this holds because $\hat{x}
\not\in L$ and so it does not have any witnesses).  By the definition
of~$\tau$, it follows that $\tau$ is not defined at $(\mbox{\it
trashbin}, \mbox{\it trashbin})$.

Condition~(c): Since $L \not\in \up$, there exists a string $x_0 \in
L$ that has at least two distinct witnesses. Fix the two smallest
witnesses, say $w_1$ and~$w_2$ with $w_1 \neq w_2$, for ``$x_0 \in
L$.''~~Let $a = \pair{x_0 , w_1}$, $b = \pair{x_0 , w_2}$, and $c =
\pair{x_0 , x_0}$ be three given arguments of~$\tilde{\tau}$.  Since
$\tilde{\tau}$ is total, each of $(a,b)$, $(b,c)$, $(a,b \tilde{\tau}
c)$, and $(a \tilde{\tau} b, c)$ is in the domain
of~$\tilde{\tau}$. However, it holds that
\[
 \tilde{\tau}(\tilde{\tau} (a,b),c) = 
\tilde{\tau}(\mbox{\it trashbin} , c) = 
\mbox{\it trashbin} \neq 
\pair{x_0 , x_0} = 
\tilde{\tau}(a,\pair{x_0 , x_0}) = 
\tilde{\tau}(a,\tilde{\tau}(b,c)).
\]
Hence, $\tilde{\tau}$ is not weakly associative or associative.~\qed

\subsection{Proof Sketch Related to Homan's Work}
\label{s:proofs-hom}

We present the proof of Theorem~\ref{thm:hom-1}.
In fact,
Theorem~\ref{thm:hom-1} 
follows immediately 
from Lemma~\ref{lem:hom-1}
below. 

\begin{lemma}
\label{lem:hom-1}
{\rm{}\cite{hom:m:aowf}}\quad
For every $n \in \nats$ and for every total, associative 
function (one-way or otherwise) 
$\sigma : \sigmastar \times \sigmastar \rightarrow \sigmastar$, 
there exists an element $z \in \sigmastar$ in the image of $\sigma$
whose preimage under $\sigma$ is of cardinality at least~$n$.
\end{lemma}

\noindent
{\bf Proof Sketch
of Lemma~\ref{lem:hom-1}.}~~Let $\sigma : \sigmastar \times
\sigmastar \rightarrow \sigmastar$ be any total, 
associative function.
For each string~$w$ in the image of~$\sigma$, define two sets
$L_w$ and $R_w$ as follows:
\begin{eqnarray*}
L_w & = & \{x \in \sigmastar \condition (x \neq w) \,\wedge\,
(\exists y \in \sigmastar)\, [\sigma(x,y) = w]\} ; \\
R_w & = & \{y \in \sigmastar \condition (y \neq w) \,\wedge\,
(\exists x \in \sigmastar)\, [\sigma(x,y) = w]\} .
\end{eqnarray*}
To prove the lemma, we will
show that for every $n \in \nats$, there exists a string
$z\in \sigmastar$ in the image of $\sigma$ 
for which at least one of the following two conditions is true:
\begin{enumerate}
\item[(1)] the set $L_z$ has cardinality at least~$n$;

\item[(2)] the set $R_z$ has cardinality at least~$n$.
\end{enumerate}

We use induction on~$n$.

For $n=1$, pick any two distinct strings $a,b\in\sigmastar$.  Since
$\sigma$ is total, $a\sigma b$ necessarily exists. Let $z=a\sigma b$.
Since $a\neq b$, either $a\neq z$ or $b\neq z$ (or both), making $z$
satisfy at least one of the conditions~(1) or~(2).

Let $n \geq 1$, and assume that there exists a string $z\in
\sigmastar$ such that at least one of conditions~(1) and~(2) holds
true for~$n$.
Assume that condition (1) holds for~$n$. 
(If condition~(2) holds for~$n$, an analogous argument works.)

We show that there is a string in $\sigmastar$ that satisfies at least
one of conditions~(1) and~(2) for $n+1$.  If the cardinality of the set $L_z$
in condition (1) is strictly greater than~$n$, we are done.
So, suppose condition~(1) holds with equality (for~$n$).  Then, there
exist $n$ pairs of strings $(x_1,y_1), \ldots, (x_n,y_n) \in
\sigmastar \times \sigmastar$ each having image $z$ under~$\sigma$ and
so that the $x_i$ are pairwise distinct and distinct from~$z$.

Choose any distinct strings $s_1,\ldots,s_{n^2+n+1}\in \sigmastar$ 
not contained in $\{x_1, \ldots, x_n, z\}$.  Since $\sigma$ is total,
for all $i$, $1\leq i \leq n^2+n+1$, there exists a string
$u_i\in\sigmastar$ such that
\[
    u_i = z \sigma s_i = (x_1 \sigma y_1) \sigma s_i = \cdots =
    (x_n \sigma y_n) \sigma s_i .
\]
Since $\sigma$ is associative, 
for all $i$, $1 \leq i \leq n^2+n+1$, we also have
\[
   u_i = z \sigma s_i = x_1 \sigma (y_1 \sigma s_i) = \cdots =
   x_n \sigma (y_n \sigma s_i) .
\]

If there exists some~$i$, $1 \leq i \leq n^2+n+1$, 
such that the corresponding string $u_i$ is not in $\{x_1,\ldots,x_n,z\}$, 
then $\{x_1,\ldots,x_n,z\} \subseteq L_{u_i}$.
Hence, this $u_i$ satisfies condition~(1) for $n+1$.
(This is the only place where we make use of the assertion 
``$(\forall j : 1 \leq j \leq n)\, [x_j \neq z]$'' 
that follows from the definition of~$L_z$.)

Otherwise, for each~$i$, $1 \leq i \leq n^2+n+1$, we have $u_i \in \{x_1,
\ldots , x_n, z\}$.  Thus, the $n^2+n+1=(n+1)n+1$ distinct pairs $(z ,
s_i)$ are mapped by $\sigma$ onto the $n+1$ strings $x_1, \ldots ,
x_n, z$.  By the pigeon-hole principle, there must exist some $\hat{z}
\in \{x_1,\ldots,x_n,z\}$
whose preimage under $\sigma$ has cardinality at least $n+1$.

We claim that $\hat{z}$ satisfies condition~(2) for $n+1$.  Let
$\hat{S}$ be the set of all~$s_i$, $1 \leq i \leq n^2+n+1$, such that
$\sigma(z , s_i) = \hat{z}$.  The above argument shows that the cardinality
of $\hat{S}$ is at least $n+1$.  Since $\hat{z} \in
\{x_1,\ldots,x_n,z\}$ and
\[
\{s_1,\ldots, s_{n^2 + n + 1}\}  \cap \{x_1, \ldots, x_n, z\} = \emptyset ,
\]
we have $\hat{z} \neq s_i$ for each $i$, $1 \leq i \leq n^2+n+1$.
Thus, $\hat{S} \subseteq R_{\hat{z}}$, which makes
$\hat{z}$ satisfy condition~(2) for $n+1$ and completes the
proof.~\qed

\bibliography{JustForSurveyBib}

\end{document}